\documentclass[
reprint,
dcolumn,
bm,
nofootinbib,
notitlepage,
superscriptaddress]{revtex4-2}
\usepackage{graphicx}
\usepackage{color}
\usepackage[latin1]{inputenc}
\usepackage[T1]{fontenc}
\usepackage{amsmath}
\usepackage{amssymb}
\usepackage{braket}
\usepackage{mathrsfs}
\usepackage{hyperref}
\usepackage{ulem}
\usepackage{upgreek}
\usepackage{pstricks-add}
\usepackage{natbib} 
\usepackage{listings}
\usepackage{tikz}
\usepackage{tcolorbox}
\usetikzlibrary{mindmap}

\definecolor{lightblue}{rgb}{0.145,0.6666,1}

\begin{document}
\title{A quantum chemistry approach to linear vibro-polaritonic IR spectra with perturbative electron-photon correlation} 

\author{Eric W. Fischer}
\email{ericwfischer.sci@posteo.de}
\affiliation{Institut f\"ur Chemie, Humboldt-Universit\"at zu Berlin, Brook-Taylor-Stra\ss{}e 2, D-12489 Berlin, Germany}
\affiliation{Institut f\"ur Chemie, Universit\"at Potsdam, Karl-Liebknecht-Stra\ss{}e 24-25, D-14476 Potsdam-Golm, Germany}

\author{Jan A. Syska}
\affiliation{Institut f\"ur Chemie, Universit\"at Potsdam, Karl-Liebknecht-Stra\ss{}e 24-25, D-14476 Potsdam-Golm, Germany}

\author{Peter Saalfrank}
\email{peter.saalfrank@uni-potsdam.de}
\affiliation{Institut f\"ur Chemie, Universit\"at Potsdam, Karl-Liebknecht-Stra\ss{}e 24-25, D-14476 Potsdam-Golm, Germany}
\affiliation{Institut f\"ur Physik und Astronomie, Universit\"at Potsdam, Karl-Liebknecht-Stra\ss e 24-25, D-14476 Potsdam-Golm, Germany}

\date{\today}

\let\newpage\relax

\begin{abstract}
In the vibrational strong coupling (VSC) regime, molecular vibrations and resonant low-frequency cavity modes form light-matter hybrid states, named vibrational polaritons, with characteristic IR spectroscopic signatures. Here, we introduce a quantum chemistry based computational scheme for linear IR spectra of vibrational polaritons in polyatomic molecules, which perturbatively accounts for nonresonant electron-photon interactions under VSC. Specifically, we formulate a cavity Born-Oppenheimer perturbation theory (CBO-PT) linear response approach, which provides an approximate but systematic description of such electron-photon correlation effects in VSC scenarios, while relying on molecular \textit{ab initio} quantum chemistry methods. We identify relevant electron-photon correlation effects at second-order of CBO-PT, which manifest as static polarizability-dependent Hessian corrections and an emerging polarizability-dependent cavity intensity component providing access to transmission spectra commonly measured in vibro-polaritonic chemistry. Illustratively, we address electron-photon correlation effects perturbatively in IR spectra of CO$_2$ and Fe(CO)$_5$ vibro-polaritonic models qualitatively in sound agreement with non-perturbative CBO linear response theory.
\end{abstract}

\let\newpage\relax
\maketitle
\newpage

\textit{Introduction.}--- Vibrational polaritons are light-matter hybrid states formed when molecular vibrational modes strongly interact with quantized modes of optical low-frequency cavities\cite{ebbesen2016} lying at the heart of the emerging field of vibro-polaritonic chemistry\cite{hirai2020,nagarajan2021}. Experimentally, vibrational polaritons exhibit characteristic spectroscopic signatures, specifically transitions to upper and lower vibro-polaritonic states, which have been probed by both linear\cite{shalabney2015a,shalabney2015b,george2015,long2015,george2016,chervy2018} and non-linear\cite{xiang2018} infrared (IR) spectroscopic techniques. 

Computationally, vibro-polaritonic IR spectra are commonly obtained from linear response theory, where two distinct routes exist: First, linear response approaches based on effective ground state Pauli-Fierz Hamiltonians, which are fully characterized by purely molecular properties such as the molecular ground state potential energy surface (PES) and dipole moment, easily accessible by means of standard quantum chemistry methods.\cite{li2020,fischer2021,lieberherr2023,gomez2023} Second, a linear response approach formulated in the cavity Born-Oppenheimer (CBO) approximation\cite{bonini2022,schnappinger2023,flick2017,flick2017cbo}, which relies on an extended electronic structure problem taking into account nonresonant interactions between electrons and low-frequency cavity modes inducing electron-photon correlation.\cite{flick2017cbo} The CBO formulation is more complete but also more involved, since both generalized cavity PES and dipole moments depend on both molecular and cavity coordinates. From a linear response perspective, both approaches are formulated in double-harmonic approximation, \textit{i.e.}, harmonic approximation for molecular and cavity modes in combination with a linear approximation for the ground state dipole moment. This is particularly beneficial for treating vibrations and spectra of large molecules or molecular ensembles quantum mechanically, since only Hessians and linearized dipole moments at equilibrium structures are required.
\begin{figure}[hbt!]
\begin{center}
\includegraphics[scale=1.0]{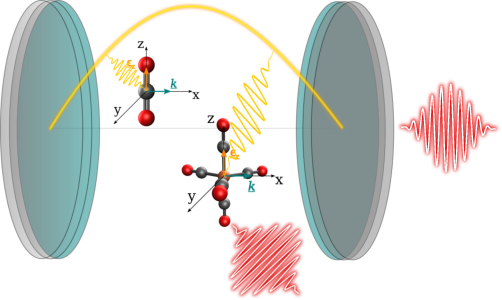}
\end{center}
\renewcommand{\baselinestretch}{1.}
\caption{Sketch of CO$_2$ and Fe(CO)$_5$ vibro-polaritonic model systems studied with respect to their linear IR response in this work.}
\label{fig.molecule_cavity}
\end{figure}

Recently, we showed that effective ground state and CBO formulations differ in the way they account for electron-photon correlation.\cite{fischer2023} Specifically, the effective ground state formulation can be understood as additional approximation to the CBO formulation, which we named \textit{crude} CBO approximation, that fully neglects electron-photon correlation in the VSC regime.\cite{fischer2023} The nonresonant nature of interactions between electrons and low-frequency cavity modes characterized by distinct excitation energy scales of both subsystems, subsequently motivated us to propose a perturbative connection between crude CBO and correlated CBO formulations, denoted as CBO perturbation theory (CBO-PT).\cite{fischer2023} CBO-PT solves the extended CBO electronic structure problem perturbatively and allows for a systematic correction of electron-photon correlation in the crude CBO approach.\cite{fischer2023} Further, CBO-PT is conceptually similar to other perturbative approaches\cite{galego2019,szidarovszky2023}, which however constitute a further approximation\cite{galego2019,fischer2023} or do not rely on the CBO formulation\cite{szidarovszky2023}.

Here, we introduce an approximation to the non-perturbative CBO linear response framework of Bonini and Flick\cite{bonini2022} by combining CBO-PT with linear response theory. This approach allows us to systematically correct vibro-polaritonic IR spectra for electron-photon correlation effects while fully relying on molecular properties, which are accessible by \textit{ab initio} quantum chemistry methods. We derive explicit expressions for electron-photon correlation corrected vibro-polaritonic Hessian matrix elements and IR intensities, which are both shown to be determined by the molecule's static polarizability, whose possible relevance has been noted only recently.\cite{szidarovszky2023,haugland2023,schaefer2023} We illustratively apply the CBO-PT linear response approach to vibro-polaritonic model systems of CO$_2$ and Fe(CO)$_5$ (\textit{cf.} Fig.\ref{fig.molecule_cavity}) in line with Ref.\cite{bonini2022}, and discuss electron-photon correlation effects in connection to experimentally relevant transmission spectra\cite{shalabney2015a,long2015}.

\textit{CBO Linear Response Theory.}--- We first recapitulate the main aspects of CBO linear-response theory\cite{bonini2022}, which fully accounts for electron-photon correlation in vibro-polaritonic IR spectra. We consider a molecular system with $N_n$ nuclei under VSC with $2N_c$ quantized transverse field modes of an IR cavity. We furthermore assume the CBO approximation to be valid, \textit{i.e.}, non-adiabatic coupling to the excited state manifold is negligible.\cite{flick2017cbo, fischer2023} The CBO electronic \textit{ground state} problem is described by an electron-photon time-independent Schr\"odinger equation (TISE)\cite{flick2017,flick2017cbo}
\begin{align}
\hat{H}_{ec}\,
\ket{\Psi^{(ec)}_0(\underline{R},\underline{x})}
&=
E^{(ec)}_0(\underline{R},\underline{x})\,
\ket{\Psi^{(ec)}_0(\underline{R},\underline{x})}
\quad,
\label{eq.electron_photon_tise}
\end{align}
with adiabatic electron-photon ground state, $\ket{\Psi^{(ec)}_0}$, and ground state cavity potential energy surface (cPES), $E^{(ec)}_0$, which parametrically depend on both nuclear, $\underline{R}$, and cavity displacement coordinates, $\underline{x}$. The electron-photon Hamiltonian reads\cite{flick2017cbo}
\begin{align}
\hat{H}_{ec}
&=
\hat{H}_e
+
V_c
+
W_c
\quad,
\label{eq.electron_photon_hamiltonian}
\end{align}
with electronic Hamiltonian, $\hat{H}_e=\hat{T}_e+V_\mathrm{coul}$, composed of the electronic kinetic energy, $\hat{T}_e$, and the molecular Coulomb potential, $V_\mathrm{coul}$. Further, $V_c=\frac{1}{2}\sum^{2N_c}_{\lambda,k}\,\omega^2_k\,x^2_{\lambda k}$, is the harmonic cavity potential characterized by harmonic frequencies, $\omega_k$, and displacement coordinates, $x_{\lambda k}$, with polarization, $\lambda$, and mode index, $k$, respectively. The third term on the right-hand side of Eq.\eqref{eq.electron_photon_hamiltonian} constitutes the light-matter interaction potential 
\begin{align}
W_c
&=
g_0
\sum^{2N_c}_{\lambda,k}
\omega_k
\left(
\underline{e}_{\lambda k}
\cdot
\underline{d}_{en}
\right)
x_{\lambda k}
+
\dfrac{g^2_0}{2}
\sum^{2N_c}_{\lambda,k}
\left(
\underline{e}_{\lambda k}
\cdot
\underline{d}_{en}
\right)^2
,
\label{eq.light_matter_potential}
\end{align}
with light-matter interaction linear in the coupling constant, $g_0$, and dipole self-energy (DSE) quadratic in $g_0$.\cite{lambdac} The coupling constant, $g_0=\frac{1}{\sqrt{\epsilon_0 V_\mathrm{cav}}}$, is only determined by the cavity volume, $V_\mathrm{cav}$, and the permittivity, $\epsilon_0$. It relates to a mode specific coupling constant, $g_k=\sqrt{2\omega_k/\hbar}\,g_0$ (\textit{cf.} Supplementary Information (SI), Sec.S1). The light-matter interaction, $W_c$, in Eq.\eqref{eq.light_matter_potential} is mediated by the polarization-projected molecular dipole operator, $\underline{e}_{\lambda k}\cdot\underline{d}_{en}$, with cavity polarization vector, $\underline{e}_{\lambda k}$. Thus, $W_c$ couples the electronic subsystem to low-frequency cavity modes and additionally induces electron-electron interactions due to the DSE term.\cite{flick2017cbo,bonini2022,fischer2023} In the non-interacting limit with $g_0=0$, the light-matter interaction, $W_c$, vanishes such that, $\hat{H}_{ec}=\hat{H}_e+V_c$, and Eq.\eqref{eq.electron_photon_tise} reduces to the \textit{electronic} TISE with a constant energy shift induced by the cavity potential, $V_c$.

In CBO linear response theory, the ground state cPES in Eq.\eqref{eq.electron_photon_tise} is harmonically approximated\cite{bonini2022}
\begin{align}
E^{(ec)}_0(\underline{R},\underline{x})
&\approx
E^{(ec)}_0(\underline{R}_0,\underline{x}_0)
+
\frac{1}{2}\,
\underline{C}^T
\underline{\underline{H}}^{(ec)}_0\,
\underline{C}
\quad,
\end{align}
around a minimum configuration, $(\underline{R}_0,\underline{x}_0)$, of the cPES with coordinate vector, $\underline{C}=(\underline{\tilde{R}},\underline{x})$, which collects $N_\mathrm{vib}$ mass-weighted cartesian displacement coordinates, $\tilde{R}_i=\sqrt{M_i}\,R_i$, with atomic mass, $M_i$, and $2N_c$ cavity displacement coordinates, $\underline{x}$. The vibro-polaritonic Hessian, $\underline{\underline{H}}^{(ec)}_0$, contains second-order derivatives of the cPES with respect to both mass-weighted molecular cartesian \textit{and} cavity displacement coordinates. It specifies a matrix eigenvalue problem\cite{bonini2022}
\begin{align}
\begin{pmatrix}
\underline{\underline{H}}^{(ec)}_{QQ} & \underline{\underline{H}}^{(ec)}_{QC} 
\vspace{0.2cm}
\\
\underline{\underline{H}}^{(ec)}_{QC} & \underline{\underline{H}}^{(ec)}_{CC}  
\end{pmatrix}
\begin{pmatrix}
\underline{q}_m
\vspace{0.3cm}
\\
\underline{c}_m 
\end{pmatrix}
=
\Omega^2_m
\begin{pmatrix}
\underline{q}_m
\vspace{0.3cm}
\\
\underline{c}_m 
\end{pmatrix}
\quad.
\label{eq.vibro_polaritonic_eigenvalue_problem}
\end{align} 
where $\underline{\underline{H}}^{(ec)}_0$ is written as $2\times2$-block matrix with molecular ($QQ$), cavity ($CC$) and light-matter interaction blocks ($QC$), respectively. Eigenvectors, $(\underline{q}_m,\underline{c}_m)^T$, resemble vibro-polaritonic normal modes with molecular, $\underline{q}_m$, and cavity components, $\underline{c}_m$, characterized by corresponding harmonic vibro-polaritonic frequencies, $\Omega_m$. The eigensystem of Eq.\eqref{eq.vibro_polaritonic_eigenvalue_problem} provides access to linear vibro-polaritonic IR spectra
\begin{align}
\sigma_\mathrm{IR}(\hbar\omega)
&=
\sum^{N_p}_m
I_m\,
\delta(
\hbar\omega
-
\hbar\Omega_m)
\quad,
\end{align}
for $N_p=N_\mathrm{vib}+2N_c$ vibro-polaritonic normal modes. IR intensities, $I_m=\sum_\kappa\vert Z_{\kappa m}\vert^2$, are determined by cartesian components of mode effective charges
\begin{align}
Z_{\kappa m}
&=
Z^{(Q)}_{\kappa m}
+
Z^{(C)}_{\kappa m}
\quad,
\quad
\kappa
=
x,y,z
\quad,
\label{eq.mode_effective_charge}
\end{align}
with molecular and cavity mode contributions\cite{bonini2022}
\begin{align}
Z^{(Q)}_{\kappa m}
&=
\sum^{N_\mathrm{vib}}_i
\dfrac{\partial D^{(ec)}_{00,\kappa}}{\partial \tilde{R}_i}\,
q_{mi}
\quad,
\vspace{0.2cm}
\\
Z^{(C)}_{\kappa m}
&=
\sum^{2N_c}_{\lambda,k}
\dfrac{\partial D^{(ec)}_{00,\kappa}}{\partial x_{\lambda k}}\,
c_{m\lambda k}
\quad,
\end{align}
where, $q_{mi}$ and $c_{m\lambda k}$, resemble elements of $\underline{q}_m$ and $\underline{c}_m$ in Eq.\eqref{eq.vibro_polaritonic_eigenvalue_problem}, respectively. The IR intensity of the $m\mathrm{th}$-vibro-polaritonic transition decomposes via Eq.\eqref{eq.mode_effective_charge} into
\begin{align}
I_m
&=
\sum_\kappa
\left[
(Z^{(Q)}_{\kappa m})^2
+
(Z^{(C)}_{\kappa m})^2
+
2
Z^{(Q)}_{\kappa m}
Z^{(C)}_{\kappa m}
\right]
\quad,
\label{eq.intensity_decomposition}
\end{align}
such that vibro-polaritonic IR spectra, $\sigma_\mathrm{IR}$, exhibit a molecular, $\sigma^{(Q)}_\mathrm{IR}$, a cavity, $\sigma^{(C)}_\mathrm{IR}$, and an additional mixed contribution, $\sigma^{(X)}_\mathrm{IR}$, respectively. Further, $Z_{\kappa m}$ is determined by cartesian components, $D^{(ec)}_{00,\kappa}$, of a generalized permanent dipole moment
\begin{align}
\underline{D}^{(ec)}_{00}(\underline{R},\underline{x})
&=
\braket{
\Psi^{(ec)}_0(\underline{R},\underline{x})
\vert
\underline{d}_{en}
\vert
\Psi^{(ec)}_0(\underline{R},\underline{x})
}_{\underline{r}}
\quad,
\label{eq.vib_pol_permanent_dipole}
\end{align}
which is evaluated with respect to the electron-photon ground state, $\ket{\Psi^{(ec)}_0}$, where integration over electronic coordinates is indicated by $\braket{\dots}_{\underline{r}}$. 

\textit{Perturbative CBO Linear Response Theory.}--- We now introduce a systematic approximation to the CBO linear response approach\cite{bonini2022} based on cavity Born-Oppenheimer perturbation theory (CBO-PT)\cite{fischer2023}. CBO-PT approximates nonresonant interactions between electrons and low-frequency cavity modes and therefore electron-photon correlation by treating the light-matter interaction potential, $W_c$, in Eq.\eqref{eq.light_matter_potential} as perturbation of the electronic subsystem\cite{fischer2023}
\begin{align}
\hat{H}_{ec}
&=
\hat{H}_0
+
\lambda\,
W_c
\quad,
\label{eq.perturbed_electron_photon_hamiltonian}
\end{align}
with zeroth-order Hamiltonian, $\hat{H}_0=\hat{H}_{e}+V_c$, and formal perturbation parameter, $\lambda$. In $\hat{H}_0$, the cavity potential, $V_c$, is a constant with respect to the bare electronic problem, such that zeroth-order states are given by bare adiabatic electronic states, $\ket{\Psi^{(e)}_\mu}$. CBO-PT is motivated by distinct excitation energy scales of electronic and low-frequency cavity mode subsystems, and provides a systematic approximation of electron-photon correlation in light-matter hybrid systems under VSC by perturbatively solving the electron-photon TISE \eqref{eq.electron_photon_tise}.\cite{fischer2023}

At $n$th-order of CBO-PT, abbreviated as CBO-PT($n$), the approximate cPES reads, $E^{(ec)}_0\approx\sum^n_{k=0}\lambda^k E^{(k)}_0$ (\textit{cf.} SI, Sec.S1), which straightforwardly leads to the corresponding Hessian
\begin{align}
\underline{\underline{H}}^{(ec)}_0
\approx
\sum^n_{k=0}
\begin{pmatrix}
\underline{\underline{H}}^{(k)}_{QQ} & \underline{\underline{H}}^{(k)}_{QC} 
\vspace{0.2cm}
\\
\underline{\underline{H}}^{(k)}_{QC} & \underline{\underline{H}}^{(k)}_{CC}  
\end{pmatrix}
\quad,
\label{eq.second_order_hessian_block}
\end{align}
here given in block-matrix form in analogy to Eq.\eqref{eq.vibro_polaritonic_eigenvalue_problem}. We restrict our discussion to perturbation order $n\leq2$, \textit{i.e.}, up to CBO-PT(2), sufficient to capture leading-order electron-photon correlation corrections\cite{fischer2023}. In contrast to the non-perturbative CBO approach\cite{bonini2022}, we work directly in a representation employing mass-weighted molecular \textit{normal modes}, which constitute the CBO-PT(0) reference and relate to coordinates, $\underline{Q}=(Q_1,\dots,Q_\mathrm{vib})$. Note, molecular normal modes are not strictly extractable from Eq.\eqref{eq.vibro_polaritonic_eigenvalue_problem} due to light-matter interaction unless $g_0=0$. 
Linear vibro-polaritonic IR spectra in CBO-PT($n$) are given by
\begin{align}
\sigma^{(n)}_\mathrm{IR}(\hbar\omega)
&=
\sum^{N_p}_m
I^{(n)}_m\,
\delta(
\hbar\omega
-
\hbar\Omega^{(n)}_m)
\quad,
\label{eq.approx_vib_pol_ir_spec}
\end{align}
with approximate vibro-polaritonic normal-mode frequencies, $\Omega^{(n)}_m$, and IR intensities, $I^{(n)}_m=\sum_\kappa\vert Z^{(n)}_{\kappa m}\vert^2$, with, $Z^{(n)}_{\kappa m}=Z^{(Q,n)}_{\kappa m}+Z^{(C,n)}_{\kappa m}$. The approximate mode effective charge, $Z^{(n)}_{\kappa m}$, is determined by components (\textit{cf.} SI, Sec.S1A for derivation)
\begin{align}
Z^{(Q,n)}_{\kappa m}
&=
\sum^{N_\mathrm{vib}}_i
\sqrt{\dfrac{\hbar}{2\omega_i}}
\dfrac{\partial D^{(n)}_{00,\kappa}}{\partial Q_i}\,
q_{mi}
\quad,
\vspace{0.2cm}
\\
Z^{(C,n)}_{\kappa m}
&=
\sum^{2N_c}_{\lambda,k}
\sqrt{\dfrac{\hbar}{2\omega_k}}
\dfrac{\partial D^{(n)}_{00,\kappa}}{\partial x_{\lambda k}}\,
c_{m\lambda k}
\quad,
\end{align}
which result from a linearization of the approximate generalized permanent dipole moment in the normal mode representation
\begin{align}
\underline{D}^{(n)}_{00}(\underline{R},\underline{x})
&=
\braket{
\Phi^{(n-1)}_0(\underline{R},\underline{x})
\vert
\underline{d}_{en}
\vert
\Phi^{(n-1)}_0(\underline{R},\underline{x})
}_{\underline{r}}
\,.
\label{eq.approx_vib_pol_permanent_dipole}
\end{align}
The latter expression is determined by the perturbatively approximated adiabatic electron-photon ground state, $\ket{\Phi^{(n-1)}_0}=\sum^{n-1}_{k=0}\lambda^k\ket{\Psi^{(k)}_0}$ (\textit{cf.} SI, Sec.S1). Note, in contrast to Eq.\eqref{eq.mode_effective_charge}, both $Z^{(Q,n)}_{\kappa m}$ and $Z^{(C,n)}_{\kappa m}$ depend on prefactors with molecular normal mode, $\omega_i$, and cavity frequencies, $\omega_k$, which follow from the normal mode representation (\textit{cf.} SI, Sec.S1A). 

We now provide explicit expressions for Hessian matrix elements and IR intensities up to CBO-PT(2) as derived in SI, Sec.S1. The sum of CBO-PT(0) and CBO-PT(1) Hessians, which \textit{do not} account for electron-photon correlation, is determined by elements (\textit{cf.} SI, Sec.S1B and S1C)
\begin{align}
\left(
\underline{\underline{H}}^{(0)}_{QQ}
+
\underline{\underline{H}}^{(1)}_{QQ}
\right)_{ij}
=&\,
\omega^2_i\,
\delta_{ij}
+
g^2_0
N_c
\sum^{2}_{\lambda}
d^{(i)}_{\lambda}
d^{(j)}_{\lambda}
\,,
\label{eq.first_order_molecular_block}
\vspace{0.2cm}
\\
\left(
\underline{\underline{H}}^{(0)}_{QC}
+
\underline{\underline{H}}^{(1)}_{QC}
\right)_{i,\lambda k}
=&\,
g_0\,
\omega_k\,
d^{(i)}_{\lambda}
\quad,
\label{eq.first_order_coupling_block}
\vspace{0.2cm}
\\
\left(
\underline{\underline{H}}^{(0)}_{CC}
+
\underline{\underline{H}}^{(1)}_{CC}
\right)_{\lambda k,\lambda^\prime k^\prime}
=&\,
\omega^2_k\,
\delta_{k k^\prime}
\delta_{\lambda\lambda^\prime}
\quad.
\label{eq.first_order_cavity_block}
\end{align} 
Here, the DSE term induces inter-normal-mode couplings in Eq.\eqref{eq.first_order_molecular_block}. Light-matter interaction elements in Eq.\eqref{eq.first_order_coupling_block} relate to polarization-projected permanent dipole moment derivatives, $d^{(i)}_{\lambda}=\partial_{Q_i}\left(\underline{e}_\lambda\cdot\underline{d}_{00}\right)\vert_{Q_i=0}$. We assume here identical polarization vectors for all modes with index, $k$. 
The CBO-PT(1) IR spectrum, $\sigma^{(1)}_\mathrm{IR}$, is fully determined by the zeroth-order adiabatic electronic ground state, $\ket{\Psi^{(e)}_0}$, such that, $\underline{D}^{(1)}_{00}=\underline{d}_{00}$ in Eq.\eqref{eq.approx_vib_pol_permanent_dipole} with first-order IR intensity given by
\begin{align}
I^{(Q,1)}_m
&=
\sum_\kappa
\vert
Z^{(Q,1)}_{\kappa m}
\vert^2
\quad,
\vspace{0.2cm}
\\
Z^{(Q,1)}_{\kappa m}
&=
\displaystyle\sum^{N_\mathrm{vib}}_i
\sqrt{\dfrac{\hbar}{2\omega_i}}
\dfrac{\partial d_{00,\kappa}}{\partial Q_i}\,
q^{(1)}_{mi}
\quad,
\end{align}
which reduces in the non-interacting limit with $g_0=0$ to $q^{(1)}_{mi}=\delta_{mi}$ and the molecular intensity, $I^{(Q,1)}_m=I^{(Q,1)}_i$. Notably, in CBO-PT(1) the vibro-polaritonic IR spectrum is fully determined by molecular intensity components, $I^{(Q,1)}_m$, since the cavity mode effective charge vanishes, $Z^{(C,1)}_{\kappa m}=0$. Thus, vibro-polaritonic contributions enter only via cavity-induced linear combinations of molecular normal mode components, $q^{(1)}_{mi}$, and the CBO-PT(1) IR spectrum, $\sigma^{(1)}_\mathrm{IR}$, exclusively probes the effective \textit{matter response} of the light-matter hybrid system under VSC. Further, CBO-PT(2) Hessian matrix element corrections in Franck-Condon approximation are given by (\textit{cf.} SI, Sec.S1D and S2) 
\begin{align}
\left(
\underline{\underline{H}}^{(2)}_{QQ}
\right)_{ij}
&=
-
\dfrac{g^4_0}{4}
N^2_c
\sum^{2}_{\lambda,\lambda^\prime}
d^{(i)}_{\lambda }
\alpha^0_{\lambda\lambda^\prime}
d^{(j)}_{\lambda^\prime }
\quad,
\label{eq.second_order_molecular_block}
\vspace{0.2cm}
\\
\left(
\underline{\underline{H}}^{(2)}_{QC}
\right)_{i,\lambda k}
&=
-
\dfrac{g^3_0}{2}
\omega_k\,
N_c
\sum^{2}_{\lambda^\prime}
\alpha^0_{\lambda\lambda^\prime}
d^{(i)}_{\lambda^\prime}
\quad,
\label{eq.second_order_coupling_block}
\vspace{0.2cm}
\\
\left(
\underline{\underline{H}}^{(2)}_{CC}
\right)_{\lambda  k ,\lambda^{\prime} k^{\prime}}
&=
-
g^2_0\,
\omega_k\,
\omega_{k^\prime}\,
\alpha^0_{\lambda\lambda^\prime}
\quad,
\label{eq.second_order_cavity_block}
\end{align}
which are all determined by polarization-projected ground state (superscript $0$) static polarizability tensor elements, $\alpha^0_{\lambda\lambda^\prime}$. Notably, the leading-order correction in Eq.\eqref{eq.second_order_cavity_block} enters the cavity block and exhibits a matter-induced perturbative coupling of distinct cavity modes\cite{fischer2023} in agreement with non-perturbative CBO linear response theory\cite{bonini2022}. 
\begin{figure*}[hbt!]
\begin{center}
\includegraphics[scale=1.0]{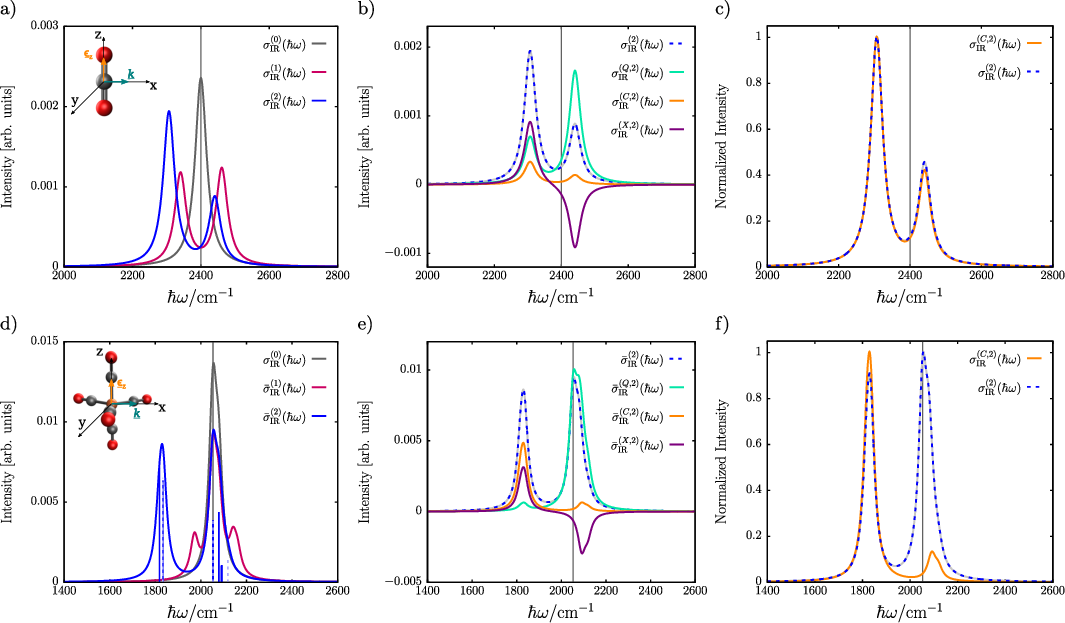}
\end{center}
\renewcommand{\baselinestretch}{1.}
\caption{Linear vibro-polaritonic IR spectra for selected single-molecule models under VSC with a single cavity mode at coupling strength, $g_0=0.03\,\sqrt{E_h}/e a_0$. Top-row: z-polarized IR spectra of the antisymmetric CO$_2$-stretching mode under VSC with a single cavity mode, $\hbar\omega_c=\hbar\omega_\mathrm{as}=2400\,\mathrm{cm}^{-1}$, with a) CBO-PT(1) and CBO-PT(2) IR spectra, $\sigma^{(1)}_\mathrm{IR}(\hbar\omega)$ and $\sigma^{(2)}_\mathrm{IR}(\hbar\omega)$, besides bare molecular spectrum, $\sigma^{(0)}_\mathrm{IR}(\hbar\omega)$, b) molecular, cavity and mixed CBO-PT(2) contributions, $\sigma^{(Q,2)}_\mathrm{IR}(\hbar\omega),\sigma^{(C,2)}_\mathrm{IR}(\hbar\omega)$ and $\sigma^{(X,2)}_\mathrm{IR}(\hbar\omega)$ and c) comparison of normalized $\sigma^{(2)}_\mathrm{IR}(\hbar\omega)$ and $\sigma^{(C,2)}_\mathrm{IR}(\hbar\omega)$. Bottom-row: Polarization-averaged linear vibro-polaritonic IR spectra for the CO-stretch band of a single Fe(CO)$_5$ molecule under VSC with a single cavity mode, $\hbar\omega_c=\hbar\omega_{e^\prime}=2052\,\mathrm{cm}^{-1}$ with d)-f) providing same information as a)-c). Stick spectra in d) resemble individual polarization-dependent contributions to $\bar{\sigma}^{(2)}_\mathrm{IR}$ (\textit{cf.} main text for details). The respective cavity mode frequency is indicated by a grey vertical line in all panels.}
\label{fig.spectra_cbopt_single}
\end{figure*}

The CBO-PT(2) IR spectrum, $\sigma^{(2)}_\mathrm{IR}(\hbar\omega)$, is determined by second-order vibro-polaritonic frequencies, $\Omega^{(2)}_m$, and intensities derived from second-order mode effective charges as
\begin{align}
I^{(2)}_m
&=
\sum_\kappa
\vert
Z^{(2)}_{\kappa m}
\vert^2
\quad,
\vspace{0.2cm}
\\
Z^{(Q,2)}_{\kappa m}
&=
\sum^{N_\mathrm{vib}}_i
\sqrt{\dfrac{\hbar}{2\omega_i}}
\dfrac{\partial D^{(Q,2)}_{00,\kappa}}{\partial Q_i}\,
q^{(2)}_{mi}
\quad,
\vspace{0.2cm}
\\
Z^{(C,2)}_{\kappa m}
&=
\sum^{2N_c}_{\lambda,k}
\sqrt{\dfrac{\hbar}{2\omega_k}}
\dfrac{\partial D^{(C,2)}_{00,\kappa}}{\partial x_{\lambda k}}\,
c^{(2)}_{m\lambda k}
\quad,
\end{align}
which contains \textit{both} molecular \textit{and} cavity contributions with dipole derivatives explicitly given by (\textit{cf.} SI Sec.S3)
\begin{align}
\dfrac{\partial D^{(Q,2)}_{00,\kappa}}{\partial Q_i}
&=
d^{(i)}_{00,\kappa}
-
\dfrac{g^2_0}{2}
N_c
\sum^{2}_{\lambda}
\alpha^0_{\kappa\lambda}\,
d^{(i)}_{\lambda}
\label{eq.molecular_dipole_component}
\quad,
\vspace{0.2cm}
\\
\dfrac{\partial D^{(C,2)}_{00,\kappa}}{\partial x_{\lambda k}}
&=
-
g_0\,\omega_k\,
\alpha^0_{\kappa\lambda}
\label{eq.cavity_dipole_component}
\quad.
\end{align}
with second-order dipole moment, $\underline{D}^{(2)}_{00}$, following from Eq.\eqref{eq.approx_vib_pol_permanent_dipole}. Note, in the non-interacting limit, $q^{(2)}_{mi}=\delta_{mi}$ and $\frac{\partial D^{(C,2)}_{00,\kappa}}{\partial x_{\lambda k}}=0$, such that the CBO-PT(2) IR spectrum reduces to the molecular spectrum. 
In Eq.\eqref{eq.molecular_dipole_component} and \eqref{eq.cavity_dipole_component}, we find a static polarization-dependent correction of the molecular intensity and in particular a non-vanishing cavity component, which is determined by the light-matter interaction strength, $g_0$, and the static polarizability, $\alpha^0_{\kappa\lambda}$, of the molecular subsystem. Notably, the related cavity dipole component, $D^{(C,2)}_{00,\kappa}$, is linear in $g_0$ and $x_{\lambda k}$ (\textit{cf.} SI, Sec.S3) in agreement with numerical results reported for non-perturbative CBO linear response theory under VSC (\textit{cf.} supporting information of Ref.\cite{bonini2022}) In contrast to CBO-PT(1), the non-vanishing CBO-PT(2) cavity intensity component allows for addressing transmission spectra, which exclusively probe the cavity response, $\sigma^{(C)}_\mathrm{IR}$, of the light-matter hybrid system commonly measured in vibro-polaritonic chemistry.\cite{shalabney2015a,long2015,george2016} In this context, we like to emphasize, that Eq.\eqref{eq.cavity_dipole_component} solely emerges due to the fact that we (perturbatively) account for electron-photon correlation effects in the description of the vibro-polaritonic system by exclusively relying on molecular ground state information. The latter is obtainable from standard quantum chemistry with polarizabilities being practically accessible by means of response theory.

\textit{IR spectra under VSC.} --- In the remainder of this work, we discuss IR spectra of CO$_2$ and Fe(CO)$_5$ vibro-polaritonic models\cite{bonini2022} obtained from second-order CBO-PT linear response theory with a focus on electron-photon correlation effects. We consider polarization-averaged vibro-polaritonic IR spectra
\begin{align}
\bar{\sigma}^{(n)}_\mathrm{IR}(\hbar\omega)
&=
\dfrac{1}{K}
\sum^K_\alpha
\sigma^{(n)}_{\alpha}(\hbar\omega)
\quad,
\quad
n=0,1,2
\quad,
\label{eq.average_ir_spec}
\end{align}
to mimic random molecular orientation, where the sum runs over all unique polarization components, $\alpha=x,y,z$, with $K=1,2,3$, depending on the number of contributing terms. Molecular structures, harmonic frequencies, dipole derivatives and static polarizability tensors elements are obtained from density functional theory (DFT) using the TPSSh\cite{tao2003,staroverov2003} and B3LYP\cite{becke1993} functionals and a Def2TZVP basis set\cite{weigend2005,weigend2006}. Computational details are provided in SI, Sec.S4A.  We furthermore consider spontaneous emission-induced Lorentzian peak broadening (and neglect other dissipative channels) with FWHM, $\kappa=41\,\mathrm{cm}^{-1}$ (\textit{cf.} SI, Sec.S4B) for an experimentally motivated cavity quality factor, $Q=\frac{\hbar\omega_c}{\kappa}$\cite{ulusoy2020,fischer2022}, with $Q=50$ (CO$_2$) and $Q=59$ (Fe(CO)$_5$).\cite{george2016}

\textit{CO$_2$ under VSC.}--- We first discuss the antisymmetric stretching mode of CO$_2$ strongly coupled to a single resonant cavity mode, $\hbar\omega_c=\hbar\omega_\mathrm{as}$, with $\hbar\omega_\mathrm{as}=2400\,\mathrm{cm}^{-1}$ (TPSSh/Def2TZVP) at coupling strength, $g_0=0.03\,\sqrt{E_h}/e a_0$. Here, the IR spectrum reduces to, $\sigma^{(n)}_\mathrm{IR}(\hbar\omega)=\sigma^{(n)}_z(\hbar\omega)$, since only the dipole component along the molecular z-axis contributes. 
In Fig.\ref{fig.spectra_cbopt_single}a, CBO-PT(1) and CBO-PT(2) IR spectra are shown for CO$_2$ under VSC besides the bare molecular spectrum, $\sigma^{(0)}_\mathrm{IR}(\hbar\omega)$. We observe a Rabi-splitting characterized by lower and upper vibro-polaritonic transitions for both CBO-PT(1) and CBO-PT(2). Uncorrelated CBO-PT(1) spectra exhibit nearly identical molecular and photonic contributions for both polariton states, contrasted by a dominantly photonic lower and dominantly molecular upper polariton state in correlated CBO-PT(2) spectra in agreement with non-perturbative CBO results\cite{bonini2022} (\textit{cf.} SI, Sec.S4B). Further, the correlated CBO-PT(2) spectrum exhibits a pronounced peak asymmetry in favour of the lower polariton transition, a spectral red-shift and a pronounced asymmetric peak splitting relative to the cavity frequency. Rabi splittings, $\Omega^{(1)}_R=121\,\mathrm{cm}^{-1}$ and $\Omega^{(2)}_R=134\,\mathrm{cm}^{-1}$, indicate a slightly stronger effective light-matter interaction in presence of electron-photon correlation, despite the fact that both results have been obtained at identical light-matter coupling. 
\begin{figure*}[hbt!]
\begin{center}
\includegraphics[scale=1.0]{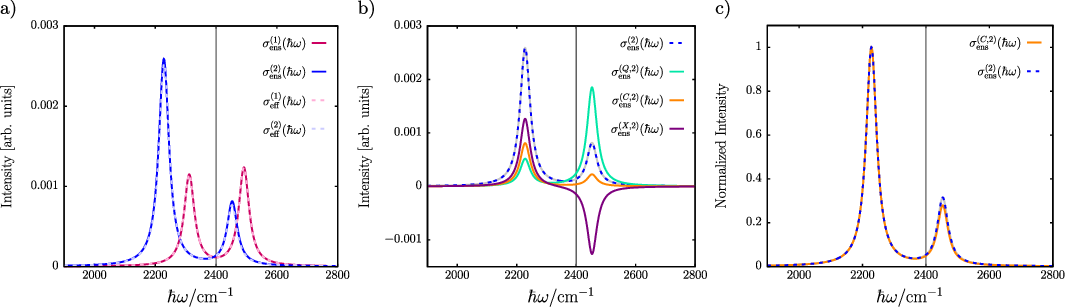}
\end{center}
\renewcommand{\baselinestretch}{1.}
\caption{Linear vibro-polaritonic IR spectra for CO$_2$ ensemble models under VSC with a single cavity mode, $\hbar\omega_c=\hbar\omega_\mathrm{as}=2400\,\mathrm{cm}^{-1}$. a) Ensemble and effective CBO-PT(n) IR spectra, $\sigma^{(n)}_\mathrm{ens}(\hbar\omega)$ and $\sigma^{(n)}_\mathrm{eff}(\hbar\omega)$ for $n=,1,2$, b) molecular, cavity and mixed CBO-PT(2) contributions, $\sigma^{(Q,2)}_\mathrm{ens}(\hbar\omega),\sigma^{(C,2)}_\mathrm{ens}(\hbar\omega)$ and $\sigma^{(X,2)}_\mathrm{ens}(\hbar\omega)$ for explicit ensemble and c) comparison of normalized $\sigma^{(2)}_\mathrm{ens}(\hbar\omega)$ and $\sigma^{(C,2)}_\mathrm{ens}(\hbar\omega)$ for explicit ensemble. The respective cavity mode frequency is indicated by a grey vertical line.}
\label{fig.spectra_cbopt_multi}
\end{figure*}

We now turn to the details of the CBO-PT(2) IR spectrum, Fig.\ref{fig.spectra_cbopt_single}b, where we compare molecular, $\sigma^{(Q,2)}_\mathrm{IR}$, cavity, $\sigma^{(C,2)}_\mathrm{IR}$, and mixed contributions, $\sigma^{(X,2)}_\mathrm{IR}$, following from Eq.\eqref{eq.intensity_decomposition}. Although, $\sigma^{(Q,2)}_\mathrm{IR}$ is dominant relative to $\sigma^{(C,2)}_\mathrm{IR}$, $\sigma^{(X,2)}_\mathrm{IR}$ determines the peak asymmetry of $\sigma^{(2)}_\mathrm{IR}$ by ``shifting'' intensity from the upper to the lower vibro-polaritonic transition. This effect directly translates to the sign of the photonic contribution in linear combinations forming upper (+) and lower (-) polariton states (\textit{cf.} SI, Sec.S4B): Mixed intensity contributions satisfy, $I^{(X,2)}_{\mp}\propto \pm\vert Z^{(Q,2)}_{\kappa \mp}\vert\vert Z^{(C,2)}_{\kappa \mp}\vert$, since the dipole derivative in Eq.\eqref{eq.cavity_dipole_component} determining $Z^{(C,2)}_{\kappa \mp}$ is negative. 
Further, we recall that IR \textit{transmission} spectra of vibro-polaritonic systems measured in micro-fluidic optical cavities probe the cavity response of the light-matter hybrid system.\cite{shalabney2015a,long2015,george2016} The cavity response translates here into the cavity component, $\sigma^{(C,2)}_\mathrm{IR}$, which appears in CBO-PT(2) but is not accounted for in uncorrelated CBO-PT(1) models\cite{bonini2022}. In Fig.\ref{fig.spectra_cbopt_single}c, we compare full and cavity CBO-PT(2) IR spectra, where the intensity has been normalized for reasons of comparison. We find nearly identical results with only minor intensity deviations for the lower polariton peak in the spectral region of interest. However, when the full IR spectrum is considered, then $\sigma^{(2)}_\mathrm{IR}$ still contains molecular contributions related to CO$_2$ bending modes at low frequencies (\textit{cf.} SI, Sec.S4B). Those peaks are absent in $\sigma^{(C,2)}_\mathrm{IR}$ since bending modes do not couple to the ``high frequency'' cavity mode. 

\textit{Fe(CO)$_5$ under VSC.}--- As a second example, we discuss the CO-stretch band of Fe(CO)$_5$ under VSC, which constitutes a both experimentally and theoretically relevant molecular multi-mode example.\cite{george2016,bonini2022,chen2022} The CO-stretch band decomposes into a doubly-degenerate equatorial normal mode (symmetry $e^\prime$ in molecular point group $D_{3h}$) and a single axial normal mode ($a^{\prime\prime}_2$ in $D_{3h}$) with harmonic frequencies, $\hbar\omega_{e^\prime}=2052\,\mathrm{cm}^{-1}$ and $\hbar\omega_{a^{\prime\prime}_2}=2078\,\mathrm{cm}^{-1}$ (TPSSh/Def2TZVP, \textit{cf}. SI, Sec.S4A). We consider a single cavity mode tuned resonant to the equatorial $e^\prime$-modes, $\hbar\omega_c=\hbar\omega_{e^\prime}$, at light-matter interaction, $g_0=0.03\,\sqrt{E_h}/ea_0$. Here, $e^\prime$-modes are $(x,z)$-polarized, while the $a^{\prime\prime}_2$-mode is $y$-polarized. Therefore, all three molecular modes can enter VSC depending on the cavity mode polarization, which constitutes a scenario with both strongly coupled and \textit{decoupled} molecular modes. In Fig.\ref{fig.spectra_cbopt_single}d, polarization-averaged CBO-PT(1) and CBO-PT(2) IR spectra are shown besides the bare molecular CBO-PT(0) spectrum. Note, the $a^{\prime\prime}_2$-peak in the latter is only visible as blue-shifted shoulder relative to the dominant transition for the herein chosen peak broadening. Further, we display CBO-PT(2) stick spectra to highlight the polarization-dependence of the individual spectral contributions. According to the averaged character, we only discuss \textit{effective} polaritonic signatures resulting from broadening effects. 
We find the $e^\prime$-peak to split under VSC but in contrast to CO$_2$, we find here additionally a pronounced molecular contribution related to \textit{uncoupled} molecular modes. Note, those \textit{do not} refer to ``dark states'' as present in the ensemble scenario below. The correlated CBO-PT(2) IR spectrum exhibits a strong red-shift and significantly more intense peaks relative to the uncorrelated CBO-PT(1) result. Both share a slight peak asymmetry in favour of the upper polariton peak. Further, a comparison of Rabi splittings, $\Omega^{(1)}_R=169\,\mathrm{cm}^{-1}$ and $\Omega^{(2)}_R=228\,\mathrm{cm}^{-1}$, indicates also here an enhanced effective light-matter interaction in presence of electron-photon correlation. By analysing, $\bar{\sigma}^{(2)}_\mathrm{IR}$, in terms of, $\bar{\sigma}^{(Q,2)}_\mathrm{IR}, \bar{\sigma}^{(C,2)}_\mathrm{IR}$ and $\bar{\sigma}^{(X,2)}_\mathrm{IR}$, reveals a trend similar to the CO$_2$ example (\textit{cf.} Fig.\ref{fig.spectra_cbopt_single}e): $\bar{\sigma}^{(Q,2)}_\mathrm{IR}$ is the dominant contribution relative to $\bar{\sigma}^{(C,2)}_\mathrm{IR}$, but the cross term again ``shifts'' intensity from the upper to the lower vibro-polaritonic transition. Importantly, in contrast to the CO$_2$ example, $\bar{\sigma}^{(Q,2)}_\mathrm{IR}$ contains here in addition to the matter response of the light-matter hybrid system bare molecular contributions, which relate to \textit{uncoupled} CO-stretching modes. In order to access the light-matter response without bare molecular contributions, we compare normalized, $\bar{\sigma}^{(2)}_\mathrm{IR}$ and $\bar{\sigma}^{(C,2)}_\mathrm{IR}$, in Fig.\ref{fig.spectra_cbopt_single}f: Here, we find the upper polariton peak to significantly differ in intensity for $\bar{\sigma}^{(C,2)}_\mathrm{IR}$, which directly relates to the absence of additional molecular contributions of uncoupled CO-modes present in $\bar{\sigma}^{(2)}_\mathrm{IR}$. Thus, correlation corrected CBO-PT(2) IR spectra allow for avoiding artificial molecular intensity contributions related to \textit{uncoupled} molecular modes in vibrational multi-mode systems under VSC, which are not probed in vibro-polaritonic transmission spectroscopy.\cite{shalabney2015a,long2015,george2016}

\textit{Collective VSC effects in CO$_2$ ensembles.}--- We finally address collective strong coupling effects in two different molecular ensemble models of CO$_2$ under VSC (\textit{cf.} SI, Sec.5), which have also been studied in Ref.\cite{bonini2022}: First, an explicit ensemble of $M$ molecules in the \textit{dilute gas} limit\cite{sidler2023}, which interact only via DSE contributions at light-matter coupling, $g_0=0.01\,\sqrt{E_h}/ea_0$. Second, an effective ``ensemble'' model, which contains a single CO$_2$ molecule under VSC with a single cavity mode characterized by an effective enhanced interaction constant, $g_\mathrm{eff}=\sqrt{M}\,g_0$. 
In Fig.\ref{fig.spectra_cbopt_multi}a, we compare ensemble, $\sigma^{(n)}_\mathrm{ens}$, and effective IRspectra, $\sigma^{(n)}_\mathrm{eff}$, for CBO-PT(1) and CBO-PT(2) for $M=20$ parallel aligned CO$_2$ molecules. Intensities have been scaled by a factor of $M^{-1}$ for reasons of comparison. We find an excellent agreement between spectra obtain at both levels of CBO-PT in agreement with Ref.\cite{bonini2022}. Further, for the ensemble model, we observe qualitatively identical results for the different spectral contributions to $\sigma^{(2)}_\mathrm{ens}$ in Fig.\ref{fig.spectra_cbopt_multi}b as in the single-molecule model discussed in Fig.\eqref{fig.spectra_cbopt_single}b. In addition, both the full IR spectrum and the bare cavity response are very similar with a minor intensity mismatch for the upper polariton peak (\textit{cf.} Fig.\ref{fig.spectra_cbopt_multi}c). We like to note here that the explicit ensemble hosts $M-1$ ``dark states'' with vanishing intensity at the bare molecular frequency, $\hbar\omega_\mathrm{as}=\hbar\omega_c$, which resemble purely molecular linear combinations of \textit{all} CO$_2$ vibrations participating in VSC and are to be contrasted by \textit{uncoupled} molecular states with \textit{non-vanishing} intensity in the IR spectrum of Fe(CO)$_5$ under VSC. From the ensemble analysis, we conclude that transmission spectra obtained from CBO-PT linear response theory can be calculated via effectively scaled single-molecule approaches subject to an effective light-matter interaction constant, $g_\mathrm{eff}=\sqrt{M}\,g_0$, when inter-molecular interactions are neglected in the ensemble in line with the non-perturbative CBO approach\cite{bonini2022}.

\textit{Conclusions.}--- We introduced a cavity Born-Oppenheimer perturbation theory (CBO-PT) linear response approach to calculate electron-photon correlation corrected vibro-polaritonic IR spectra of polyatomic molecules under VSC. This approach approximates non-perturbative CBO linear response theory\cite{bonini2022} by perturbatively approximating nonresonant interactions between electrons and low-frequency cavity modes, while fully relying on molecular \textit{ab initio} quantum chemistry methods. In zeroth-order, we obtain the bare molecular spectrum and in first-order the linear vibro-polaritonic IR spectrum in absence of electron-photon correlation. Electron-photon correlation is accounted for at second-order of CBO-PT linear response theory, where it manifests as static polarizability-dependent cavity intensity component and related corrections of Hessian matrix elements, most notably a matter-mediated inter-cavity mode coupling. Those corrections capture the main characteristics of non-perturbative CBO linear response theory\cite{bonini2022} and provide additional physical insight due to their explicit relation to molecular and cavity mode properties. A comparison of uncorrelated CBO-PT(1) and correlation corrected CBO-PT(2) IR spectra for CO$_2$ and Fe(CO)$_5$ vibro-polaritonic models reveals the impact of electron-photon correlation on vibro-polaritonic intensity ratios, Rabi splittings and experimentally relevant transmission spectra, which probe only the cavity response of the light-matter hybrid system. In both model scenarios, CBO-PT(2) linear response theory exhibits significant qualitative agreement with non-perturative CBO linear response results reported in Ref.\cite{bonini2022}. In summary, CBO-PT linear response theory constitutes a promising since easily accessible approach to vibro-polaritonic IR spectra of polyatomic molecular systems under VSC accounting for non-trivial electron-photon correlation effects. A systematic quantitative comparison of CBO-PT and non-perturbative CBO linear response theories based on controllable \textit{ab initio} wave function approaches\cite{schnappinger2023} will be subject of future work.

\section*{Acknowledgements}
The authors acknowledge funding by the Deutsche Forschungsgemeinschaft (DFG, German Research Foundation) under Germany's Excellence Strategy within ``Sonderforschungsbereich 1636'' of University Potsdam, project A05, ``Understanding and controlling reactivity under vibrational and electronic strong coupling''. E.W. Fischer acknowledges the kind hospitality of Michael Roemelt and his group at Humboldt-Universit\"at zu Berlin, and helpful discussions with Thomas Schnappinger (Stockholm). J.A. Syska acknowledges support and funding by the International Max Planck Research School for Elementary Processes in Physical Chemistry.

\section*{Data Availability Statement}
The data that support the findings of this study are available from the corresponding author upon reasonable request.

\section*{Conflict of Interest}
The authors have no conflicts to disclose.



\begin{thebibliography}{200}
\bibitem{lambdac} The interaction parameter $g_0$ is equivalent to $\lambda_c$ employed by Bonini and Flick.
\end{thebibliography}
\end{document}